\documentclass[prl,floatfix,twocolumn,showpacs,superscriptaddress]{revtex4}
\usepackage{graphicx}
\def\bea{\begin{eqnarray}}
\def\eea{\end{eqnarray}}
\def\be{\begin{equation}}
\def\ee{\end{equation}}

\def\S{\mbox{\bf S}}

\begin{document}

\author{Didier Poilblanc}

\affiliation{
  Laboratoire de Physique Th\'eorique, Universit\'e Paul Sabatier,
  F-31062 Toulouse, France }
\affiliation{
Institute of Theoretical Physics,
Ecole Polytechnique F\'ed\'erale de Lausanne,
BSP 720,
CH-1015 Lausanne,
Switzerland}

\author{Andreas L\"auchli}
\affiliation{
  Institut Romand de Recherche Num\'erique en Physique des Mat\'eriaux (IRRMA),
  PPH-Ecublens, CH-1015 Lausanne}

\author{Matthieu Mambrini}

\affiliation{
  Laboratoire de Physique Th\'eorique, Universit\'e Paul Sabatier,
  F-31062 Toulouse, France }

\author{Fr\'ed\'eric Mila}
\affiliation{
Institute of Theoretical Physics,
Ecole Polytechnique F\'ed\'erale de Lausanne,
BSP 720,
CH-1015 Lausanne,
Switzerland}

\date{\today}
\title{Spinon deconfinement in doped frustrated quantum antiferromagnets}

\begin{abstract}
The confinement of a spinon liberated by doping two-dimensional
frustrated quantum antiferromagnets with a non-magnetic impurity
or a mobile hole is investigated. For a static vacancy, an
intermediate behavior between complete deconfinement (kagome) and
strong confinement (checkerboard) is identified in the
$J_1{-}J_2{-}J_3$ model on the square lattice, with the emergence
of two length scales, a spinon confinement length {\it larger}
than the magnetic correlation length. For mobile holes, this
translates into an extended spinon-holon boundstate  allowing one
to bridge momentum (ARPES spectral function) and real space (STM)
experimental observations. These features provide clear evidence
for a nearby "deconfined critical point" in a doped microscopic
model.
\end{abstract}
\maketitle

The search for exotic spin liquids (SL) has been enormously amplified
after the discovery of the high critical temperature (high-$T_C$)
cuprate superconductors.
Indeed, Anderson suggested that the Resonating Valence Bond state
is the relevant insulating parent state that would become immediately
superconducting under hole doping~\cite{RVB}.
Such a state is characterized by
short range magnetic correlations and no continuous (spin) or discrete
(lattice) broken symmetry. Another major characteristic
of this SL phase is the deconfinement~\cite{spinons} of the
S=1/2 excitations (spinons) in contrast to ordered magnets which
have $S=1$ spin waves.
Upon doping, some scenarii predict a 2D
Luttinger liquid~\cite{2D-LL}, i.e. a state which exhibits
spin-charge separation, a feature generic of one-dimensional
correlated conductors.

Magnetic frustration is believed to be the major tool
to drive a two-dimensional (2D) quantum antiferromagnet
(AF) into exotic quantum disordered phases. The Valence Bond Solid
(VBS), an alternative class of quantum disordered phases breaking
lattice symmetry, seems to be a strong candidate in some
frustrated quantum magnets as suggested by robust field
theoretical arguments~\cite{VBS}, early numerical computations of
frustrated quantum AF on the square lattice with diagonal
bonds~\cite{j1j2,plaquette,plaquette_note} and
in the 2D checkerboard lattice~\cite{checkerboard} (with diagonal
bonds only on half of the plaquettes). In contrast, the 2D kagome
lattice~\cite{KagomeED} shows no sign of ordering
of any kind~\cite{KagomeED_2,KagomeSinglets}.
The "deconfined critical point"
(DCP), a new class of quantum criticality, was proposed recently
e.g. to describe a {\it continuous} AF to VBS transition~\cite{Senthil}.

\begin{figure}
\centerline{\includegraphics*[angle=0,width=0.9\linewidth]{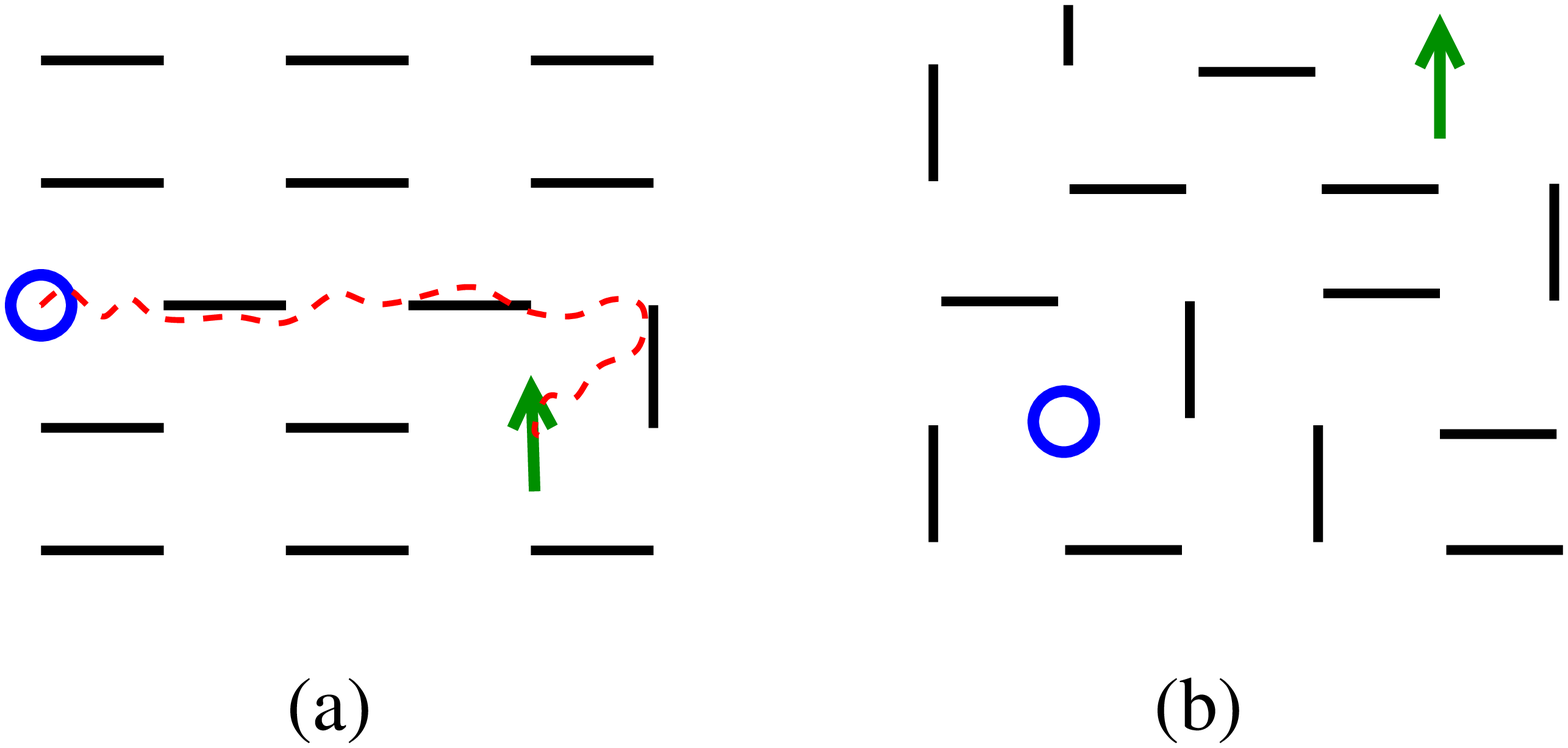}}
  \caption{\label{fig:confine}
(Color on-line) Schematic picture of a vacancy (or
doped hole) in a frustrated magnet.
The segments stand for singlet bonds and the
arrow represents the spinon liberated in the process.
(a) Holon-spinon BS in a columnar VBS bound by a ``string'' potential
(dotted line).
(b) Deconfined holon and spinon in an (hypothetical) SL host.}
\end{figure}

Investigation of hole doping in frustrated
magnets~\cite{PRL1} has revealed stricking differences
between VBS and SL phases although they both exhibit
a finite spin-spin correlation length $\xi_{\rm AF}$.
Viewing these phases
as fluctuating singlet backgrounds,
removing an electron at a given site (e.g. by chemical substitution
with an inert atom) or, as in
Angular Resolved Photoemission Spectroscopy (ARPES)
experiments, in a Bloch state of given
momentum naturally breaks a spin dimer and liberates a spinon,
i.e. a S=1/2 polarisation in the vicinity of the empty site (holon).
The single hole spectral function shows a sharp
peak (resp. a broad feature) characteristic of a holon-spinon boundstate
(resp. holon-spinon scattering states) in the checkerboard VBS
phase (resp. kagome
SL phase).
The simple physical pictures behind these two typical
behaviors are depicted in Fig.~\ref{fig:confine}(a) and (b) for
a confining columnar dimer phase and a SL phase respectively.
The new length scale $\xi_{\rm conf}$ (average distance
between vacancy and spinon) which emerges
naturally in the VBS phase is to be identified
with the correlation length over which dimer (or VBS) order sets in.
Interestingly, it has been predicted that,
in the vicinity of the DCP,
confinement occurs on a much larger length scale
$\xi_{\rm conf}$ which diverges as
a power law of $\xi_{\rm AF}$~\cite{Senthil}.

\begin{figure}
\centerline{\includegraphics*[angle=0,width=0.9\linewidth]{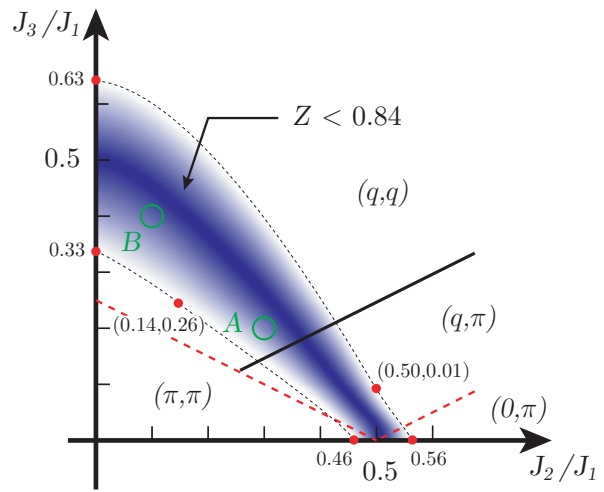}}
  \caption{\label{fig:phase_diag}
(Color on-line) Classical phase diagram for the $J_1$-$J_2$-$J_3$
model. Second order (discontinuous) transitions are indicated by
dashed (solid) lines (see e.g. Ref.~\protect\cite{Ferrer}).
The shaded (blue online) region shows the approximate location of the
minimum of the spectral weight $Z$ in the quantum version.
The region with a weight between $0.79$ and $0.84$ on the 32-site
cluster is delimited by dashed lines and red dots.
}
\end{figure}

So far the DCP scenario is only supported
by field theoretic arguments. It is therefore of crutial importance to
investigate its relevance in the framework of microscopic models.
In this Letter, we address the issues depicted in Fig.~\ref{fig:confine}
by considering a single hole introduced in
the 2D spin-1/2 AF $J_1$-$J_2$-$J_3$ Heisenberg model on the square lattice
 at zero temperature defined by
\begin{equation}
   H= \sum_{\langle ij\rangle}J_{ij}\S_i \cdot \S_j
\label{eq:H}
\end{equation}
where the $J_{ij}$ exchange parameters
are limited to first ($J_1$), second ($J_2$)
and third ($J_3$) N.N. AF couplings.
The classical phase diagram of this model~\cite{J1J2J3,Ferrer} is
very rich (see Fig.~\ref{fig:phase_diag})
showing four ordered states -- N\'eel,
collinear (${\bf q}=(\pi,0)$)
and two helicoidal -- separated by continuous or discontinuous boundaries.
The subtle interplay between quantum fluctuations and frustration ($J_2$ and
$J_3$ terms) is
expected to destabilize the classical phases and lead to a quantum disordered
singlet ground state, possibly of VBS type.
We show that one of the major prediction of the DCP scenario, namely the
emergence of a hierarchy of length scales is indeed observed for
intermediate frustration in correlation with the possibility
of a direct N\'eel-VBS continuous transition. This finding is contrasted to
two other extreme behaviors -complete deconfinement and strong confinement-
observed in the kagome and checkerboard lattices respectively.

Let us first briefly review some results in the literature
supporting the existence of a cristaline quantum disordered
phase in model~(\ref{eq:H}) (leading to spinon confinement).
In the parameter
range where frustration is largest,
many approaches, including spin-wave theory~\cite{ChandraDoucot},
exact diagonalizations~\cite{j1j2},
series-expansion~\cite{GelfandSinghHuse} and large-$N$
expansions~\cite{spinons}, have firmly established
for  $J_3=0$ the relative stability of a quantum disordered
singlet ground state:
a columnar valence bond solid with both translational and rotational
broken symmetries~\cite{VBS} or a plaquette state with no broken rotational
symmetry~\cite{plaquette}
have been proposed.
For the pure
$J_1-J_3$ model, a non-classical phase also appears between the
N\'eel $(\pi,\pi)$ and the spiral $(q,q)$ phases~: a VBS columnar
state~\cite{Leung}
or a succession
of a VBS and $Z_2$ spin-liquid phases~\cite{CapriottiSachdev} have
been proposed. Lastly, when $J_2$ and $J_3$ are both non-zero,
in the range $(J_3+J_2)/J_1 \sim 0.4-0.6$,
the finite size scaling analysis of
the dimer susceptibility computed up to 50 sites~\cite{Mambrini} shows a
non-vanishing signal
again strongly suggesting a VBS
order~\cite{plaquette_note}.

A static vacancy at a given
site O of the lattice is a simple setup relevant
to test the DCP ideas and to experiments.
In pratice, the vacancy is simply modelled by setting to zero
all the couplings $J_{ij}$ involving site O and the computations
are performed by Lanczos ED of a
cluster of 32 sites (i.e. $\sqrt{32}\times\sqrt{32}$)
which respects all point group symmetries of the infinite lattice.
Such an impurity acts, theoretically, as a local probe of the host.
It can be viewed alternatively as a localized holon ($S=0$ and charge $Q=e$)
so that the form of the surrounding spin density is expected to
provide valuable insights on the spin-charge confinement/deconfinement
mechanism.

\begin{figure}
\centerline{\includegraphics*[angle=-90,width=0.9\linewidth]{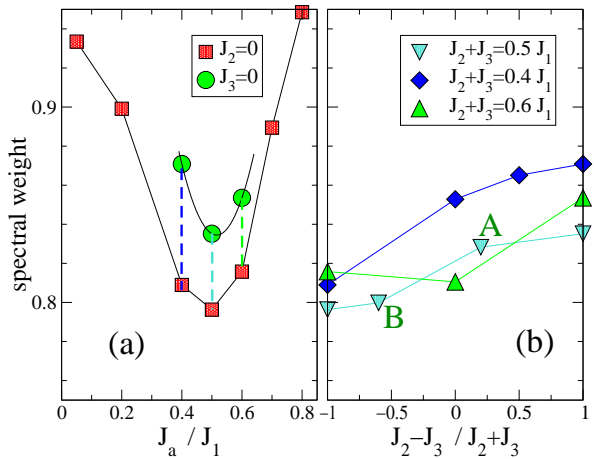}}
  \caption{\label{fig:weight_static}
(Color on-line) Static hole (vacancy) spectral weight vs AF
exchange parameters.
(a) vs $J_2/J_1$ for $J_3=0$ and vs $J_3/J_1$ (for $J_2=0$)
as indicated on plot.
(b) Along three different lines $(J_2+J_3)/J_1={\rm cst}$
in the 2D ($J_2/J_1$,$J_3/J_1$) parameter space. A and B
refer to the points in the phase diagram of Fig.~\protect\ref{fig:phase_diag}.}
\end{figure}

The single impurity Green function $G(\omega)=\big<\Psi_{\rm
bare}|(\omega-H)^{-1}|\Psi_{\rm bare}\big>$ is computed by (i)
constructing the (normalized) ''bare'' initial state $|\Psi_{\rm
bare}\big>=2c_{O,\sigma}|\Psi_{0}\big>$ from the host GS
$|\Psi_{0}\big>$ by removing an electron of spin $\sigma$ and (ii)
using a standard Lanczos continued-fraction technique. Most of the
$\omega$-integrated spectral weight (normalized to 1) of ${\rm
Im}G(\omega)$ is in fact contained in the lowest energy pole of
weight $Z=|\big<\Psi_{\rm gs}|\Psi_{\rm bare}\big>|^2$ where
$|\Psi_{\rm gs}\big>$ is the (normalized) GS of the system with
one vacancy at site $O$. Results shown in
Fig.~\ref{fig:weight_static}(a,b) show however that $Z$ is
significantly suppressed in the region where a quantum disordered
state is expected. We show in Fig.~\ref{fig:phase_diag} the region
corresponding to a reduced weight on the 32-site
cluster.

\begin{figure}
\centerline{\includegraphics*[angle=0,width=0.9\linewidth]{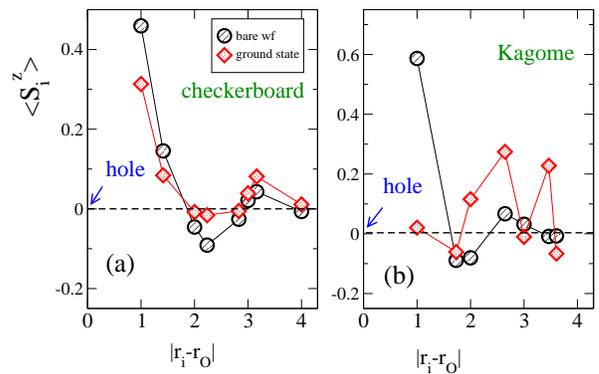}}
\caption{\label{fig:siz1} (Color on-line) Spin polarization in the
vicinity of the vacancy (summed up on equivalent sites) for both
the ''bare'' vacancy state
 and the GS for
 (a) the checkerboard lattice (32 site cluster) and (b) the kagome lattice (30 site
cluster). }
\end{figure}

\begin{figure}
\centerline{\includegraphics*[angle=0,width=0.9\linewidth]{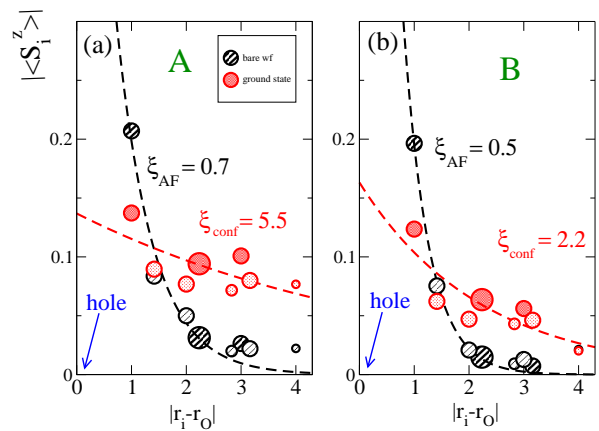}}
\caption{\label{fig:siz2} (Color on-line) Same as
Fig.~\protect\ref{fig:siz2} (but for the {\it modulus})
for the $J_1$-$J_2$-$J_3$ model with
$J_2/J_1=0.3$ and $J_3/J_1=0.2$ (a) and $J_2/J_1=0.1$ and
$J_3/J_1=0.4$ (b) corresponding to points A
and B
in the phase diagram of
Fig.~\protect\ref{fig:phase_diag}. Fits using exponential forms
are shown in dashed lines. The areas of the dots are proportional
to the number of equivalent sites from the vacancy (entering in
the fits). Dark and light symbols correspond to positive and negative values
respectively.
  }
\end{figure}

The reduction of $Z$ is the first signal that the
spinon {\it moves away} from the original site next to the
vacancy at an average distance $\xi_{\rm conf}$ to be determined.
A quantitative measure of this effect is provided by a careful
inspection of the average local spin density $\big<S_i^z\big>$
around the vacancy in both the ''bare'' wavefunction and the true
GS.  Note that $\big<S_i^z\big>$ in $|\Psi_{\rm bare}\big>$ gives
the initial spin-spin correlation $\big<S_O^zS_i^z\big>_0$
in the host GS. We start this analysis by examining the two extreme
behaviors provided by the Heisenberg model on the checkerboard and
the kagome lattices reported in Fig.~\ref{fig:siz1}(a) and (b)
respectively. Clearly, the results for the checkerboard lattice
show very short-ranged and incommensurate spin-spin correlations.
In addition the spinon remains almost entirely confined on the
N.N. site of the vacancy. In contrast, on the kagome lattice, the
spin-1/2 delocalizes on the whole lattice, a clear signature of
deconfinement. Results for the J$_1$-J$_2$-J$_3$ model in
Fig.~\ref{fig:siz2}(a,b) for parameters corresponding to the two
typical A and B points of the phase diagram of
Fig.~\ref{fig:phase_diag} (chosen because of a reduced $Z$ factor)
reveal completely new behaviors. First,
we observe for both $A$ and $B$ very short magnetic correlation
lengths characterized by a fast oscillating decay (with the AF
wavevector) of the correlations (see below). Note that no sign of
incommensurability is seen in the oscillations of
Fig.~\ref{fig:siz2}(a) unlike in the classical spiral phase.
Interestingly, the behavior of $\big<S_i^z\big>$ in the
''relaxed'' $|\Psi_{\rm gs}\big>$ state differs drastically from
the bare state with {\it a much slower decay with
distance}~\cite{Bulut}. As seen in Fig.~\ref{fig:siz2}, accurate
fits can be realized by assuming a simple exponential decay
together with an oscillatory behavior at wavevector
$(\pi,\pi)$~\cite{note}. The very short correlation length
$\xi_{\rm AF}$, below one lattice spacing, is to be contrasted
with the strikingly large {\it confinement} length $\xi_{\rm conf}$
typically ranging from 2 to 6 lattice spacings~\cite{note2}.

Let us now discuss some of the important implications of such
findings. First, we note that such non-trivial extended spin
structure could be seen experimentally. Indeed, the substitution
of a S=1/2 atom by a non-magnetic one (e.g. Zn$^{2+}$ for
Cu$^{2+}$) which acts as a vacant site can be exactly described by
our previous model. Moreover, the local spin densities
$\big<S_i^z\big>$ on the magnetic sites around a vacancy (spinless
atom) {\it in the bulk} can be directly accessed by Nuclear
Magnetic Resonance (NMR). It is important to notice that NMR would
probe the spinon ''wavefunction'' in the ''relaxed'' state and not
the host spin correlations. Note that newly developed
spin-polarized Scanning Tunnelling Microscopy (SP-STM) techniques
might also allow to probe such atomic-scale spin
structure~\cite{SP-STM} around a vacancy {\it on a surface}.
Secondly, at the theoretical level, the numerical evidence for a
clear hierarchy of length scales, $\xi_{\rm conf}
\gg \xi_{\rm AF}$ provides a strong argument in favor of
the new class of DCP~\cite{Senthil}. Incidently,
as seen in Fig.~\ref{fig:phase_diag}, it is quite plausible that
A and B lie indeed in the vicinity
of the (supposed) N\'eel to VBS phase transition
line, the paradigm of the DCP.

\begin{figure}
\centerline{\includegraphics*[angle=0,width=0.75\linewidth]{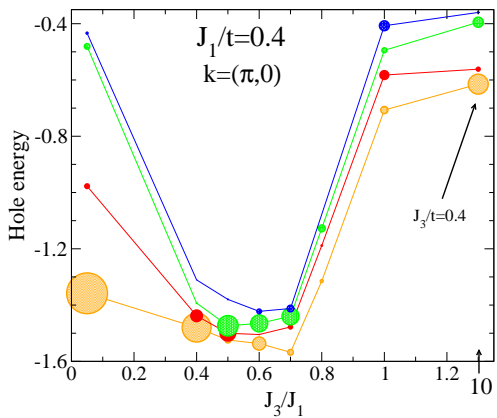}}
  \caption{\label{fig:Akw}
(Color on-line)
Four lowest energy poles of the
single hole Green function (for hole momentum ${\bf
k}=(\pi,0)$) vs increasing frustration $J_3/J_1$. The GS
energy of the undoped AF sets the energy reference. We assume here
$J_2=0$ and a fixed ratio of $J_1/t=0.4$ ($J_3/t=0.4$) for $J_3\le
J_1$ ($J_3>J_1$). The areas of dots are proportional to the spectral
weights $|\big< \Psi^{(n)}_{\rm 1 hole}|\Psi_{\rm
bare}\big>|^2$, $n=1,...,4$. To set the scale, $Z\simeq 0.04$ ($n=1$)
for $J_3=0.6$.}
\end{figure}

Lastly, we examine the case of a mobile hole. This
mimics an ARPES experiment in a Mott
insulator where a single photo-induced hole is created or the case
of a small chemical doping. The hole motion
described as in a t--J model is characterized by a hole hopping
amplitude $t$. For the unfrustrated t--J model, the hole dynamics
has been successfully analyzed in term of holon-spinon
boundstate~\cite{string}. Note that adding motion to the hole
charge leads alone to a
large reduction of $Z$, e.g. from 0.93 ($t=0$) down to 0.36 ($J_1/t=0.4$)
at small frustration ($J_3/J_1=0.05$).
As seen in Fig.~\ref{fig:Akw}, as frustration is increased, (i) the
low-energy spectral weight decreases further and (ii) the
quasiparticle peak (at the bottom of the spectrum)
is rapidly redistributed on several poles.
This remarkable behavior indicates a severe weakening of the
binding between the two constituents or, equivalently, a rapid
increase of the size of the holon-spinon boundstate.
A spectral weight $Z$ below $0.01$ is typical in this
maximally frustrated region (e.g. $Z\simeq 0.008$ for $J_3/J_1=0.5$).
The dynamic hole problem being in fact smoothly
connected to the case of
a static hole~\cite{Sachdev}, our data at finite $t$ bring
then additional strong evidence for the proximity of a DCP.

To conclude, the confinement of a spinon liberated by introducing
a vacant site or a mobile hole has been studied in various
frustrated Heisenberg AF. In the region of large frustration
of the $J_1{-}J_2{-}J_3$ model, an intermediate behavior between a strong
confinement (as in the checkerboard Heisenberg model) and a
complete deconfinement (as
on the kagome lattice) is observed, suggesting the emergence of a
new length scale related to the confinement of the spinon. Its
large value compared to the spin-spin correlation length
supports the field-theoretic
"deconfined critical point" scenario~\cite{Senthil} for the
N\'eel-VBS transition.
Furthermore, an interesting connection between this
real-space picture and features in the hole spectral function is
established.

We thank IDRIS (Orsay, France) for allocation of CPU-time on the
NEC-SX5 supercomputer. D.P. acknowledges the Institute for Theoretical Physics
(EPFL, Switzerland) for partial support.
This work was supported by the Swiss National Fund
and by MaNEP.


\begin{thebibliography}{}
\bibitem{note} The incommensurability seen
in Fig.~\protect\ref{fig:siz2}(b) at long distances
does not affect the fit accuracy.
\end{thebibliography}
\end{document}